\documentclass{JHEP3}
\usepackage{amssymb,amsmath}

\DeclareMathOperator{\tr}{Tr}
\newcommand{\Hodge}{\mathop{\star}}
\newcommand{\Real}{\mathbb{R}}
\newcommand{\e}{\textrm{e}}
\newcommand{\de}{\textrm{d}}
\newcommand{\im}{\mathrm{i}}
\newcommand{\SO}{\mathop{\mathrm{SO}}}
\newcommand{\SL}{\mathop{\mathrm{SL}}}
\newcommand{\GL}{\mathop{\mathrm{GL}}}
\newcommand{\cN}{\mathcal{N}}
\newcommand{\QE}{Q_\mathrm{e}}
\newcommand{\QM}{Q_\mathrm{m}}

\author{Jan Perz\\Afdeling Theoretische Fysica,
Katholieke Universiteit Leuven\\
Celestijnenlaan 200D bus 2415, 3001 Heverlee, Belgium\\
E-mail: \email{Jan.Perz@fys.kuleuven.be}}
\author{Paul Smyth\\II. Institut f\"ur Theoretische Physik der
Universit\"at Hamburg\\Luruper Chaussee 149, 22761 Hamburg, Germany\\
E-mail: \email{paul.smyth@desy.de}}
\author{Thomas Van Riet\\
Departamento de F\'isica Te\'orica y del Cosmos\\
and Centro Andaluz de F\'isica de Part\'iculas Elementales\\
Universidad de Granada, 18071 Granada, Spain\\ and \\
Departamento de F\'isica, Universidad de Oviedo,\\
Avda.~Calvo Sotelo 18, 33007 Oviedo, Spain\\
E-mail: \email{thomasvr@itf.fys.kuleuven.be}}
\author{Bert Vercnocke\\Afdeling Theoretische Fysica,
Katholieke Universiteit Leuven\\
Celestijnenlaan 200D bus 2415, 3001 Heverlee, Belgium\\
E-mail: \email{Bert.Vercnocke@fys.kuleuven.be}}

\title{First-order flow equations for extremal and non-extremal black
holes}

\abstract{%
We derive a general form of first-order flow equations for extremal
and non-extremal, static, spherically symmetric black holes in
theories with massless scalars and vectors coupled to gravity. By
rewriting the action as a sum of squares \`a la Bogomol'nyi, we
identify the function governing  the first-order gradient flow, the
`generalised superpotential', which reduces to the `fake
superpotential' for non-supersymmetric extremal black holes and to the
central charge for supersymmetric black holes. For theories whose
scalar manifold is a symmetric space after a timelike dimensional
reduction, we present the condition for the existence of a generalised
superpotential. We provide examples to illustrate the formalism in
four and five spacetime dimensions. }

\keywords{black holes in string theory, black holes, integrable
equations in physics}

\preprint{CAFPE-108/08, KUL-TF-08/26, UG-FT-238/08}

\begin{document}

\section{Introduction}

A distinctive feature of supersymmetric extremal black holes with
regular event horizons in theories of gravity coupled to neutral
scalar fields and Abelian vector fields is the attractor mechanism
\cite{Ferrara:1995ih,Ferrara:1996dd,Ferrara:1996um,Ferrara:1997tw}.
Its name derives from the fact that the radial evolution of the
scalars follows a set of first-order equations, such that near the
event horizon the scalars are driven to values determined by the
electric and magnetic charges carried by the black hole. For
supersymmetric black holes these equations are implied by
supersymmetry and constitute a gradient flow on the scalar manifold,
governed by the central charge. The attractor mechanism also applies
to some extremal non-supersymmetric black holes. This suggests the
possibility of non-supersymmetric gradient flows. Indeed, fake
superpotentials have been found for some non-supersymmetric extremal
black holes \cite{Ceresole:2007wx,Andrianopoli:2007gt,Cardoso:2007ky}.

Conversely, it has been shown that the existence of first-order flow
equations is not necessarily tied to the attractor mechanism, since
there exist non-extremal solutions described by such flow equations
\cite{Miller:2006ay} and non-extremal black holes cannot be attractive
(see e.g.~\cite{Garousi:2007zb}). This raises the question of whether
one can find a general form of the flow equations which is valid for
both extremal and non-extremal black holes, and under what conditions
these equations constitute a gradient flow. Having a first-order
description at hand for non-extremal solutions might shed light on
some open problems concerning the relation between the scalar charges
and the entropy of non-extremal black holes \cite{Ferrara:2008hw}.
Furthermore, a fake superpotential is the natural candidate for a
$c$-function for non-BPS solutions
\cite{Goldstein:2005rr,Andrianopoli:2007gt}.

In this paper we present the general form of the gradient flow
equations valid for extremal and non-extremal, static and spherically
symmetric solutions, extending the formalism developed in
\cite{Ceresole:2007wx} for extremal solutions. We name the function
that determines the gradient flow the `generalised superpotential' in
analogy with the fake supergravity formalism for domain walls
\cite{Freedman:2003ax,Celi:2004st,Skenderis:2006jq,
Papadimitriou:2006dr,Afonso:2006gi,Elvang:2007ba, Sonner:2007cp}. In
addition, for theories whose moduli space is a symmetric space after a
timelike dimensional reduction, we derive the condition for a
generalised superpotential to exist. In these cases the black hole
equations of motion are explicitly integrable
\cite{Breitenlohner:1987dg,Gaiotto:2007ag,Gunaydin:2005mx,
Bergshoeff:2008be}. In fact, in the case of extremal, but not
necessarily supersymmetric black holes with regular horizons, the
procedure proposed in \cite{Andrianopoli:2007gt} should be sufficient
to construct a fake superpotential for symmetric moduli spaces in
$\cN$-extended supergravities, as long as the fake superpotential can
be expressed in terms of duality invariants. However, for general
extremal and non-extremal solutions (including those without regular
horizons) not much is known. The goal of our work is to fill this gap,
providing a uniform description of extremal and non-extremal black
holes.

We begin our discussion by  recalling the necessary background
material (section \ref{s:Prerequisites}). In particular, we briefly
recall what is known about the construction of a black hole effective
action and first-order flow equations from the existing literature. We
show in section \ref{s:FlowFormalism} how one can obtain such a
one-dimensional effective action with a black hole potential, in an
arbitrary number of spacetime dimensions, and how to find the most
general first order flow equations from a `generalised
superpotential', assuming that it exists. This is illustrated by an
example, the dilatonic black hole, in section \ref{s:Illustration}.
Section \ref{s:Existence} discusses the question of the existence of a
superpotential in arbitrary dimensions. In section
\ref{s:FreeParticle}, we explain how to obtain a free-geodesic form of
the effective action by timelike dimensional reduction, for systems
whose scalar manifold is a symmetric space after the reduction, and
derive from it first-order equations. In sections \ref{single} and
\ref{multiple} we then study this condition for a single-scalar and a
multi-scalar example. We end with a discussion of our results and a
comparison with the literature on domain walls in section
\ref{s:discussion}.

\section{Prerequisites}\label{s:Prerequisites}

\subsection{Two forms of the black hole effective action}

We will consider static, spherically symmetric black hole solutions in
gravity coupled to a number of neutral scalars $\phi^a$ and vector
fields $A^I$ in $D+1$ dimensions,
\begin{equation}\label{E-M-D-action}
S = \int \de^{D+1}x \sqrt{|g|}\Bigl(\mathcal{R}_{D+1} -
\tfrac{1}{2}G_{ab}\partial_\mu\phi^a\partial^\mu\phi^b -
\tfrac{1}{2}\mu_{IJ} F^I_{\mu\nu} F^{J\,\mu\nu}\Bigr),
\end{equation}
where $G_{ab}$ and $\mu_{IJ}$ are functions that depend on the scalar
fields $\phi ^ a$, and $F^I_{\mu\nu}$ are Abelian field strengths.
Greek indices are raised and lowered with the spacetime metric
$g_{\mu\nu}$ and $g=\det g_{\mu\nu}$. For now we leave the dimension
unspecified, but note that in the special case $D+1=4$ there can be
another term in the action of the form, $-\tfrac12
\nu_{IJ}F^I_{\mu\nu}(\Hodge F^J)^{\mu\nu}$, where $\nu_{IJ}$ also
depends on the scalar fields. To keep our discussion as general as
possible we shall make no further assumptions about this theory, but
the reader should notice that it is of the appropriate form to
describe the bosonic sector of ungauged supergravity.

There are two techniques to construct the effective action for such
systems. Both are based on the fact that static, spherically symmetric
solutions depend only on the radial parameter, so that effectively the
problem is one-dimensional. The first technique \cite{Gibbons:1996af,
Ferrara:1997tw} expresses the Maxwell field strengths in terms of the
magnetic and electric charges (the fluxes of $F$ and $\Hodge F$ at
spatial infinity) via the respective equations of motion (and Bianchi
identities). Consider for example the metric ansatz for a black hole
in $D+1 = 4$ dimensions
\begin{equation}\label{m1}
\de s^2 = -\e^{2U(\tau)}\de t^2
+ \e^{-2U(\tau)}\gamma_{mn}\de x^m\de x^n\,,
\end{equation}
where $U(\tau)$ is often referred to as the black hole warp factor and
depends only on the radial coordinate $\tau$ on the spherically
symmetric spatial slice with the metric $\gamma_{mn}$. The
one-dimensional effective action obtained as explained above turns out
to be that of a particle subject to an external force field given by
the effective black hole potential $V$:
\begin{equation}
S=\int \de \tau \,\Bigl( 2\dot{U}^2
+\tfrac{1}{2}G_{ab}(\phi)\dot{\phi}^a\dot{\phi}^b +
\e^{2U}V(\phi)\Bigr),\label{action1}
\end{equation}
where a dot means differentiation with respect to the radial parameter
$\tau$. The configuration space of this `fiducial' particle is a
direct product $\mathcal{M}\times \Real$ where $\mathcal{M}$ is the
scalar target space, with metric $G_{ab}$, and $\Real$ represents the
warp factor. The `mass parameters' in the black hole potential $V$ are
given by the electric and magnetic charges. Solutions to this action
have to obey a constraint, stemming from part of the information in
the $D+1$-dimensional Einstein equations that cannot be derived from
the effective action.\footnote{This constraint can be found from the
effective action if one introduces an `einbein' corresponding to the
reparametrisations of the radial coordinate. This einbein then acts as
a Lagrange multiplier that enforces the constraint
\cite{Janssen:2007rc}.} In section \ref{s:FlowFormalism} we will
explain how to use this first method in arbitrary $D+1$ dimensions.

The second technique for constructing a one-dimensional effective
action, first described in the $D+1 = 4$ case in
\cite{Breitenlohner:1987dg}, is based on the observation that a static
solution in $D+1$ dimensions can be dimensionally reduced over time to
a Euclidean $D$-dimensional instanton solution. Because of the assumed
spherical symmetry, the resulting instanton solutions are carried only
by the metric and the scalars in $D$ dimensions. Moreover, since the
reduction is performed over a Killing direction, the $D$-dimensional
solutions fully specify the solutions in $D+1$ dimensions. As
explained in \cite{Breitenlohner:1987dg, Bergshoeff:2008be} the
equations for the $D$-dimensional metric decouple and are easily
solved. The scalar field equations of motion are found from the
following effective one-dimensional action
\begin{equation}
S=\int \de \tau \,\tilde G_{ij}\dot{\tilde\phi}^i\dot{\tilde\phi}^j\,,
\label{action2}
\end{equation}
which describes the free geodesic motion of a fiducial particle in an
enlarged target space of scalar fields $\tilde\phi^i$ that contain the
scalar fields $\phi^a$ of the $(D+1)$-dimensional theory plus
axion-type scalar fields arising from the reduced vector potentials.
In the remainder of this paper, we will always use the notation
$\tilde G$ for the moduli space metric in the reduced (Euclidean)
gravity theory. Note that in this procedure the vectors (or
equivalently, the axions) are not eliminated by their equations of
motion. It is these axionic scalars that have the opposite sign for
their kinetic term, which causes the metric $\tilde{G}$ to have an
indefinite signature. If we were to start in four dimensions and would
then integrate out those axions, we would find the other black hole
effective action (\ref{action1}). This second technique will be
explored in section \ref{s:FreeParticle}. There we will discuss
systems for which the moduli space after reduction to $D$ dimensions
is a symmetric space and show how to extract the $D+1$ dimensional
first order equations.

\subsection{Flow equations}

The question of which technique (or effective action) is best suited
for the given task depends on the theory one considers and on which
aspects of black hole solutions one wishes to investigate. For
instance, if the scalar target space in the effective action of the
second type (\ref{action2}) is a symmetric space then the geodesic
equations are manifestly integrable and can be used to construct
explicit solutions, see for instance \cite{Bergshoeff:2008be} for more
details. When one is interested in studying supersymmetry and the
black hole attractor mechanism, the first approach is more commonly
used. For a supersymmetric (BPS) black hole ansatz the first-order
Killing spinor equations in $D+1$ dimensions provide an integrated
form of the second order equations of motion derived from
(\ref{action1}), and are of the type
\begin{equation}
\dot{\phi}^a=\pm G^{ab}\partial_b|Z|\,,\label{Floweq}
\end{equation}
where the function $Z$ has the property that\footnote{Note that in our
conventions the gravitational coupling constant $\kappa^2$ in the
Einstein--Hilbert term $\tfrac{1}{2\kappa^2}\sqrt{|g|}\mathcal{R}$ is
set to $\tfrac{1}{2}$. This influences the coefficients in formula
\eqref{centralcharge}.}
\begin{equation}\label{centralcharge}
|Z|^2 + 2\tfrac{(D-1)}{(D-2)}G^{ab}\partial_a |Z|\partial_b |Z|=V\,.
\end{equation}
and (when evaluated at infinity) is the (complex) central charge. The
set of equations (\ref{Floweq}) is called BPS or \emph{gradient flow
equations}, and describes an \emph{attractor flow} if there is an
attractive fixed point (that is, when the black hole potential has a
minimum). For non-supersymmetric black holes the first-order equations
are no longer guaranteed to exist. Nonetheless, as non-supersymmetric
\emph{extremal} black holes can still exhibit attractor behaviour
\cite{Goldstein:2005hq,Andrianopoli:2006ub,DAuria:2007ev,
Andrianopoli:2007kz}, it would seem plausible that they admit a
first-order description. Indeed, Ceresole and Dall'Agata
\cite{Ceresole:2007wx} have shown that it is possible for extremal
non-supersymmetric black holes to mimic the BPS equations
\eqref{Floweq} of their supersymmetric counterparts, with the central
charge replaced by a suitable `superpotential' function $W \neq \lvert
Z\rvert$ (which is not necessarily a remnant of supersymmetry in one
dimension higher \cite{Cardoso:2007ky}). Subsequent work has provided
further examples of the hidden structure in non-supersymmetric
extremal solutions, see e.g.~\cite{Andrianopoli:2007gt,Ferrara:2007qx,
Ferrara:2008hw,Ferrara:2008ap,Bellucci:2008sv} and references therein.
Of most direct relevance for this work is \cite{Andrianopoli:2007gt},
where a fake superpotential was presented for $\cN>2$ theories in
$D+1=4$, which all possess a symmetric moduli space after timelike
reduction.

While non-extremal black holes are of considerable interest, little is
known about their possible interpretation as solutions of first-order
equations. As non-extremal solutions cannot be attractors even when a
regular horizon exists (see \cite{Garousi:2007zb}), it is perhaps
already surprising that some non-extremal solutions can be found from
first-order equations derived from a superpotential \cite{Lu:2003iv}.
Miller et al.~\cite{Miller:2006ay} later provided the simplest
possible example---the non-extremal Reissner-Nordstr\"om black
hole---by making use of Bogomol'nyi's trick from non-gravitational
field theories, namely completing the squares.\footnote{The
Bogomol'nyi trick was first applied to self-gravitating solutions in
the case of cosmic strings \cite{Comtet:1987wi}; see also
\cite{Collinucci:2006sp,Davis:2008ps} for recent discussions. The same
procedure can be applied to time-dependent gravitating solutions
\cite{Chemissany:2007fg}.} In these theories one is able to rewrite
the energy functional as a strict sum of squares. The
energy-minimising solutions are found by solving the first-order
Bogomol'nyi equations that result from setting each of the squares to
zero, and correspond to the BPS solutions in the supersymmetric
completion of the original field theory. It was pointed out in
\cite{Miller:2006ay} that the coupling to gravity introduces at least
one term with a relative minus sign, which would appear to ruin this
scheme. However, one can show that the extremal, static, BPS solutions
can be found by solving the equivalent set of first-order equations
that arise in rewriting the total action in terms of squared
expressions. It transpires that the relative minus sign makes the
rewriting of the action as a sum of squares non-unique and allows one
to introduce a one-parameter deformation. This leads to the
non-extremal version of first-order equations, with the deformation
parameter measuring the deviation from extremality.

The Bogomol'nyi approach has been generalised to include the
non-extremal dilatonic black hole and $p$-brane solutions, as well as
time-dependent (cosmological) solutions in arbitrary dimensions
\cite{Janssen:2007rc} and non-extremal black holes in gauged
supergravity \cite{Lu:2003iv, Cardoso:2008gm}. The first-order
formalism for time-dependent solutions is of interest as it provides
further evidence for hidden structures in cosmologies, as first
suggested by the domain wall/cosmology correspondence
\cite{Skenderis:2006jq}. The explicit structure of non-extremal flow
equations in theories with more complicated scalar matter coupling is
not known, although some suggestions were made in
\cite{Andrianopoli:2007gt}.

\section{General flow equations in $D+1$ Dimensions}
\label{s:FlowFormalism}

Consider the metric describing static, spherically symmetric black
hole solutions of the theory described by the action
(\ref{E-M-D-action}). The most general form of the spacetime metric
consistent with these symmetries is
\begin{equation}
\de s^2_{D+1} = - \e^{2\beta\varphi(\tau)}\de t^2
+ \e^{2\alpha\varphi(\tau)} \Bigl(\e^{2(D-1)A(\tau)}\de\tau^2
+ \e^{2A(\tau)}\de\Omega_{D-1}^2\Bigr),\label{metric}
\end{equation}
where
\begin{equation}
\alpha=-1/\sqrt{2(D-1)(D-2)}\,,\qquad \beta=-(D-2)\alpha\,,
\end{equation}
and the scalars depend solely on the radial coordinate: $\phi^a =
\phi^a(\tau)$. In four and five dimensions, a common notation for the
black hole warp factor is $U = -\alpha \varphi$, i.e.~in the
four-dimensional case ($D+1=4$) $U=\varphi/2$, while in five
dimensions ($D+1=5$) $U= \varphi/\sqrt{3}$.  Following the procedure
of \cite{Gibbons:1996af, Ferrara:1997tw}, described beneath
(\ref{m1}), we eliminate the vector fields in terms of the charges
through their equations of motion and obtain a one-dimensional action
of the form
\begin{equation}\label{a1}
S=\int\de\tau\,\,\Bigl( \frac{\dot{A}^2 + \e^{2(D-2)A}}{2\alpha^2}
-\tfrac{1}{2}\dot{\varphi}^2 -
\tfrac{1}{2}G_{ab}\dot{\phi}^a\dot{\phi}^b -
\e^{2\beta\varphi}V(\phi^a)\Bigr),
\end{equation}
where a dot denotes a derivative with respect to $\tau$. We use small
Latin indices from the beginning of the alphabet $a,b,\ldots$ to label
the scalars of the $(D+1)$-dimensional theory and $G$ denotes the
moduli space metric in the same theory. This action is supplemented
with a Hamiltonian constraint, which states that the radial evolution
of the fields happens on a slice of constant total energy
\begin{equation}
(2\alpha^2)^{-1}(\dot{A}^2 -\e^{2(D-2)A})
= \tfrac12\dot{\varphi}^2 + \tfrac12 G_{ab}\dot{\phi}^a \dot{\phi}^b
- \e^{2\beta\varphi}V(\phi)\equiv E\,.\label{HamiltonianI}
\end{equation}
The constraint is the remnant of the original $D+1$-dimensional
Einstein equations that is not reproduced by the effective action
(\ref{a1}). The total gravitational energy $E$ contains a charge
contribution, such that extremal black holes have vanishing energy
($E=0$) and non-extremal black holes have positive energy ($E>0$).

\paragraph{Generalised superpotential.}

Let us now assume that there exists a function $Y(\varphi,\phi^a)$,
which we call the `generalised superpotential', such that
\begin{equation}
e^{2\beta\varphi}V(\phi^a)
=\tfrac{1}{2}\partial_{\varphi} Y\partial_{\varphi} Y
+\tfrac{1}{2}\partial_a Y\partial^a Y + \Delta\,,\label{Y-equation}
\end{equation}
where $\Delta$ is a constant to be determined later (see
eq.~\eqref{eq:gammadeltaE}). The effective action (\ref{a1}) can then
be written in the following form\footnote{In fact a minus sign is also
possible within the squares, but this choice amounts to a redefinition
of $\tau$ and $Y$, so without loss of generality we may choose plus.}
\begin{equation}
S=\frac{1}{2}\int\de\tau\,\Bigl[\frac{1}{\alpha^2}
\Bigl(\dot{A}+\sqrt{\e^{2(D-2)A}+\gamma^2}\Bigr)^2
- (\dot{\varphi}+\partial_\varphi Y)^2
- (\dot{\phi}^a+\partial^a Y)^2\Bigr],\label{eq:SumSquares1}
\end{equation}
plus a total derivative; $\gamma$ is a constant.

The first-order form of the equations of motion is then obtained by
putting the terms within brackets in (\ref{eq:SumSquares1}) to zero,
giving a stationary point of the action. We first note that the
solution to the first-order equation for $A$ is independent of the
details of the model under consideration:
\begin{equation}
\e^{-(D-2)A}=\gamma^{-1}\sinh[(D-2)\gamma\tau+\delta]\,,
\end{equation}
where $\delta$ is an integration constant. The constant $\gamma^2$
appearing under the square root must be non-negative to ensure the
absence of naked singularities.  We then call the remaining equations
\emph{generalised flow equations}
\begin{align}
\dot{\varphi}+\partial_{\varphi}Y&=0\label{FLOWII}\,,\\
\dot{\phi}^a + G^{ab}\partial_{b}Y&=0\label{FLOWIII}\,.
\end{align}
The Hamiltonian constraint (\ref{HamiltonianI}) fixes the constant
$\Delta$ appearing in (\ref{Y-equation}) to be
\begin{equation}\label{eq:gammadeltaE}
(D-1)(D-2)\gamma^2=-\Delta =E\,.
\end{equation}

\paragraph{Extremal case.}

When $\Delta=0$ (extremality) equation (\ref{Y-equation}) implies that
$Y(\varphi, \phi^a)$ must factor as
\begin{equation} Y(\varphi,
\phi^a)=\e^{\beta\varphi}W(\phi^a)\,,\label{factorization}
\end{equation}
such that the formula for the black hole potential assumes the
familiar form
\begin{equation}
V = \tfrac{1}{2}\beta^2 W^2 + \tfrac{1}{2}\partial_a W\partial^a W\,,
\end{equation}
and the flow equations become the known expressions for extremal black
hole solutions
\begin{align}
\dot{\varphi}+\beta\e^{\beta\varphi}W&=0\label{FLOW'II}\,,\\
\dot{\phi}^a + \e^{\beta\varphi}\partial^{a}W&=0\label{FLOW'III}\,.
\end{align}
This means that the main difference between the flow equations
describing extremal and non-extremal solutions is the factorisation
property (\ref{factorization}) of the generalised superpotential
$Y(\varphi,\phi^a)$.

The form of the flow equations for non-extremal solutions presented
here differs somewhat from the conjecture made in
\cite{Andrianopoli:2007gt}, which proposes to preserve the form of the
flow equations from the extremal case (\ref{FLOW'II}, \ref{FLOW'III}),
but allow $W$ to explicitly depend on $\tau$: $W(\phi,\tau)$. Noting
that explicit $\tau$-dependence can locally be rephrased as
$\varphi$-dependence, with $\tau$ then considered as a function of
$\varphi$, one sees that this is in a similar vein as our proposal.
The two are not equivalent, however, as in \cite{Andrianopoli:2007gt}
the dependence of $W$ on $\tau$ is of a specific kind,
$\partial_{\tau}W\sim-\gamma^2\e^{-\varphi/2}$. With this form of
$\partial_{\tau} W$ the two sets of equations (\ref{FLOWII},
\ref{FLOWIII}) and (\ref{FLOW'II}, \ref{FLOW'III}) can hold
simultaneously only in the extremal case. In section
\ref{s:Illustration} we give an explicit example where all
non-extremal solutions obey equations (\ref{FLOWII}, \ref{FLOWIII}),
but not (\ref{FLOW'II}, \ref{FLOW'III}).

\section{An illustration: dilatonic black hole}\label{s:Illustration}

The simplest theory involving scalar fields that admits charged black
hole solutions is given by the Einstein-dilaton-Maxwell action
\begin{equation}
S=\int \de^4 x\sqrt{|g|}\Bigl(
\mathcal{R}-\tfrac{1}{2}(\partial\phi)^2
-\tfrac{1}{4}\e^{a\phi}F^2\Bigr)\,,\label{eq:EMDaction}
\end{equation}
where the dilaton coupling $a$ is a non-zero constant. In reference
\cite{Janssen:2007rc} the `fake' BPS equations for the purely
electric, extremal and non-extremal solutions of this theory were
given by writing the action as a sum and difference of squares,
generalising the results on the pure Reissner--Nordstr\"om black hole
in \cite{Miller:2006ay}. In the following we reconsider these results
in the language of section \ref{s:FlowFormalism} and extend to the
full dyonic solution. For dyonic solutions, however, we will notice
that only in the $a=1$ case can we easily find the fake
superpotential. In section \ref{single} we return to this subject and
discuss the $a=\sqrt{3}$ example. We refer the reader to
\cite{Gibbons:1982ih} for the original treatment of dilatonic black
hole solutions.

Following the language of section \ref{s:FlowFormalism}, we will now
consider the first-order equations and the construction of a
generalized superpotential for the dilaton $\phi$ and the `warp
factor' $\varphi$ appearing in the metric (\ref{metric}). Note that
for $D+1=4$ we have $\beta=-\alpha=1/2$. As explained above, the
equations of motion for $\varphi$ and $\phi$ can be derived from a
one-dimensional action of the form (\ref{a1}), where now
$G_{ab}\dot\phi^a\dot\phi^b=\dot \phi^2$ and the black hole effective
potential is given by
\begin{equation}
V(\phi) = \tfrac12 \QE^2\e^{-a\phi} + \tfrac12 \QM^2\e^{+a\phi}\,,
\end{equation}
where $\QE$ is the electric charge and $\QM$ is the magnetic charge
(which, in what follows, we assume to be non-negative).

\subsection{Purely electric or magnetic solutions}

The first-order equations found in \cite{Janssen:2007rc} for purely
electric solutions are\footnote{We changed the sign of $\beta_3$ and
divided it by $2$, compared to the definition in
\cite{Janssen:2007rc}.}
\begin{align}
f^{\varphi}(\varphi,\phi)&\equiv
\dot{\varphi}=-\tfrac{2}{1+a^2}\sqrt{\tfrac{1+a^2}{4}\QE^2\e^{
\varphi-a\phi}+\beta_2^2}- \tfrac{2a}{1+a^2}\beta_3\,,\\
f^{\phi}(\varphi,\phi)&\equiv\dot{\phi}
=+\tfrac{2a}{1+a^2}
\sqrt{\tfrac{1+a^2}{4}\QE^2\e^{\varphi-a\phi}+\beta_2^2}
-\tfrac{2}{1+a^2}\beta_3\,,
\end{align}
where we introduced a set of integration constants
($\gamma,\beta_2,\beta_3$) that obey the Hamiltonian constraint
\begin{equation}
(1+a^2)\gamma^2=\beta_2^2+\beta_3^2\,.
\end{equation}

In order for a generalised superpotential $Y$ to exist, the above
two-dimensional flow must be a gradient flow. This is locally the
case, as one immediately verifies that the curl,
$\partial_{[\phi}f_{\varphi]}$, vanishes.\footnote{In this example no
distinction needs to be made between lower and upper indices, but we
maintain it for consistency with section \ref{s:Existence}.} It is not
difficult to construct the generalised superpotential explicitly,
\begin{equation}
Y(\varphi,\phi)=-\frac{2}{1+a^2}\Bigl(2\surd{s_\mathrm{e}}
- 2\beta_2\log(\beta_2+\surd{s_\mathrm{e}}) + \beta_2(\varphi-a\phi)
+ \beta_3(a\varphi + \phi)\Bigr)\,,\label{eq:ElDilBHSupPot}
\end{equation}
where $\surd{s_\mathrm{e}}$ is shorthand for
$\sqrt{\tfrac{1+a^2}{4}\QE^2\e^{\varphi-a\phi}+\beta_2^2}$. As
expected, extremality ($\gamma=\beta_2=\beta_3=0$) implies that the
superpotential $Y$ factorises according to (\ref{factorization})
\begin{equation}
Y(\varphi,\phi)
=\e^{\varphi/2}\bigl(-\tfrac{2}{\sqrt{1+a^2}}\QE\e^{-a\phi/2}\bigr)
\equiv\e^{\varphi/2}W(\phi)\,.
\end{equation}

We now come back to the remark made at the end of the previous
section. If we compare with the flow equations of
\cite{Andrianopoli:2007gt}, by locally inverting $\varphi(\tau)$, we
do not find the form of $\partial_\tau W(\phi,\varphi(\tau))$
suggested in \cite{Andrianopoli:2007gt}, unless $\gamma = \beta_2 =
\beta_3 = 0$. However, from the expression (\ref{eq:ElDilBHSupPot})
for the generalised superpotential, we see that we can explicitly
construct $Y$ for all possible values of the nonextremality parameters
$\beta_2, \beta_3$.

In the case of purely magnetic charge, the above equations hold when
the following electromagnetic duality rule is imposed:
\begin{equation}
\QE\rightarrow \QM\,,\qquad\phi\rightarrow -\phi\,,\qquad
\beta_2\leftrightarrow\beta_3\,.
\end{equation}

\subsection{Dyonic solutions}

The dyonic case with arbitrary dilaton coupling $a$ is more involved.
The theory with $a=1$ is the simplest and, in the \emph{extremal
case}, it is not difficult to see that the correct vector field $f$
can be found by summing the electric and magnetic ones
\begin{align}
f^{\varphi}(\varphi,\phi) &=
-\tfrac{1}{\sqrt{2}}\QE\e^{(\varphi-\phi)/2}-
\tfrac{1}{\sqrt{2}}\QM\e^{(\varphi+\phi)/2}
\,,\\
f^{\phi}(\varphi,\phi) &=
+\tfrac{1}{\sqrt{2}}\QE\e^{(\varphi-\phi)/2}
-\tfrac{1}{\sqrt{2}}\QM\e^{(\varphi+\phi)/2}\,.
\end{align}
The corresponding superpotential $Y$ is
\begin{equation}
Y(\varphi,\phi)= -\e^{\varphi/2}\sqrt{2}\left(\QE\e^{-\phi/2}
+ \QM\e^{+\phi/2}\right)\equiv \e^{\varphi/2} W(\phi),
\end{equation}
and is the sum of the pure electric and magnetic superpotentials. An
extremum of the superpotential $W(\phi)$, and consequently of the
black hole potential $V(\phi)$, only exists in the dyonic case,
corresponding to the fact that an attractive $AdS_2$ horizon exists
only in the extremal dyonic case.

Let us now extend to \emph{non-extremal solutions} using the
technique of \cite{Miller:2006ay}, as explained in the previous
section. This gives
\begin{align}
f^{\varphi}(\varphi,\phi)
&=-\sqrt{\tfrac{1}{2}\QE^2\e^{\varphi-\phi}+\beta_2^2}
-\sqrt{\tfrac{1}{2}\QM^2\e^{\varphi+\phi}+\beta_3^2}\,,\\
f^{\phi}(\varphi,\phi)
&=+\sqrt{\tfrac{1}{2}\QE^2\e^{\varphi-\phi}+\beta_2^2}
-\sqrt{\tfrac{1}{2}\QM^2\e^{\varphi+\phi}+\beta_3^2} \,.
\end{align}
The corresponding generalised superpotential $Y$ reads
\begin{equation}
\begin{split}
Y(\varphi,\phi)={}&-2\surd{s_\mathrm{e}}
+2\beta_2\log(\beta_2+\surd{s_\mathrm{e}})-\beta_2(\varphi-\phi)\\
&-2\surd{s_\mathrm{m}}
+2\beta_3\log(\beta_3+\surd{s_\mathrm{m}})-\beta_3(\varphi+\phi)\,,
\end{split}
\end{equation}
where $\surd{s_\mathrm{e}}$ is defined as in the electric case and
$\surd{s_\mathrm{m}}$ is shorthand for
$\sqrt{\tfrac{1}{2}\QM^2\e^{\varphi+\phi}+\beta_3^2}$.

We have not been able to integrate the second-order equations for
$\varphi$ and $\phi$ when $a\neq 1$. However, we demonstrate in
section \ref{single} that the case $a=\sqrt{3}$ can also be solved
explicitly with the aid of the group-theoretical methods of section
\ref{s:FreeParticle}. For general dilaton coupling $a$ we are not
aware of whether the solution for the dyonic case is explicitly known
or not, but in the next section we argue that the \emph{extremal}
solution (if it exists) obeys first-order flow equations.

\section{Existence of a generalised superpotential}\label{s:Existence}

In this section we comment on the question of whether a generalised
superpotential exists or not. First, we consider black holes with a
single scalar field and explain that, at least in the extremal case, a
generalised superpotential always exists, as argued in a different way
already in \cite{Ceresole:2007wx}. We then investigate the multiscalar
case and see that a generalised superpotential exists when the
velocity field on the enlargement of the scalar manifold in $D+1$
dimensions with the warp factor $\varphi$ is irrotational (curl-free),
generalising the condition of \cite{Ceresole:2007wx} for extremal
black holes to non-extremal ones. In the following section we will
study this velocity field in detail for theories with a symmetric
moduli space after a timelike dimensional reduction.

\subsection{Black holes with a single scalar field}

The argument for the existence of a fake superpotential for extremal
black hole solutions involving one scalar field is taken from the fake
supergravity formalism for single scalar domain walls
\cite{Skenderis:2006jq} and proceeds as follows. Assume that the
extremal dyonic solution exists, then equation (\ref{FLOW'II}) can be
used to give $W$ in terms of the radial parameter $\tau$, i.e.~this
defines the function $W(\tau)$. Since the black hole is supported by a
single scalar $\phi$, we have that $W$ depends only on $\phi$. Locally
we can always invert the function $\phi(\tau)$ to $\tau(\phi)$ and
this defines $W(\phi)$.

Having constructed the fake superpotential $W(\phi)$ for the extremal
solution, we could then attempt the deformation technique of
\cite{Miller:2006ay} to obtain the function $Y(\varphi,\phi)$ in the
non-extremal case. This approach, however, requires the Lagrangian to
satisfy certain conditions (see \cite{Miller:2006ay} for details). For
the dyonic example of the previous section it turns out that only the
Lagrangian with $a=1$ obeys these constraints. Therefore, when $a\neq
1$, even though flow equations might exist, the procedure cannot be
applied. For $a=\sqrt{3}$ the hidden symmetries of the theory will
allow us to demonstrate the existence of generalised flow equations
also in the non-extremal case (section \ref{single}).

\subsection{Black holes with multiple scalar fields}
\label{s:MultiScalar}

When a black hole solution is carried by multiple scalars, the above
argument for the existence of \emph{extremal} flow equations does not
apply.\footnote{Unless some complicated conditions are satisfied, as
explained in the case of domain walls in \cite{Celi:2004st,
Sonner:2007cp}.} Furthermore, for domain walls an example has been
found, where a solution does not admit a first-order flow that can be
derived from a fake superpotential \cite{Sonner:2007cp}.

We shall now reconsider the question of the existence of a gradient
flow for black holes. Remember that, in the formalism of section
\ref{s:FlowFormalism}, the gradient of the generalised superpotential
determines the first order derivatives of both the `warp factor'
$\varphi$  appearing in the $D+1$-dimensional metric and the
$D+1$-dimensional scalars $\phi^a$, through equations (\ref{FLOWII})
and (\ref{FLOWIII}). Therefore we consider all these scalars on the
same footing and will combine them in a vector $\phi^A$:
\begin{equation}
\phi^A=\{\varphi,\phi^a\}\,.
\end{equation}
In the following section we will investigate a class of theories that
have a symmetric moduli space when reduced over one dimension, as
their equations of motion are known to be integrable. Using the
integrability of the effective action we can explicitly write down the
velocity vector field $f$ on the enlarged scalar manifold in $D$
dimensions
\begin{align}
&\dot{\phi}^A\equiv f^A(\phi,\chi)\,,\\
&\dot{\chi^{\alpha}}\equiv f^{\alpha}(\phi,\chi)\,,
\end{align}
where the $\chi^{\alpha}$ are the scalars descending from the vector
potentials upon dimensional reduction. One can demonstrate that there
are enough `integrals of motion' to fully eliminate the
$\chi^{\alpha}$ in terms of the $\phi^A$, such that one can write down
a velocity field on the original target space in $D+1$ dimensions:
\begin{equation}\label{velocity2}
\dot{\phi}^A=f^A(\phi,\chi(\phi))\,.
\end{equation}

Having obtained the velocity field (\ref{velocity2}) on the moduli
space in $D+1$ dimensions, it suffices to show that the velocity
one-form $f_A$ is locally exact
\begin{equation}\label{curl}
f_A(\phi,\chi(\phi))\equiv \tilde G_{AB}(\phi)f^B(\phi,\chi(\phi))
=\partial_A Y(\phi)\,,
\end{equation}
where $\tilde G$ is the metric on the scalar manifold in the
$D$-dimensional theory. A necessary and sufficient condition for this
to hold locally is, by Poincar\'e's lemma, that the one-form is closed
\begin{equation}\label{curl1}
\partial_{[A} f_{B]}=0\,.
\end{equation}
Whether or not the field $Y(\phi)$ is defined over the whole target
space is of less relevance to us and depends on the cohomology of the
target space.

For specific non-supersymmetric solutions it might be very difficult
in practice to find the superpotential $Y$. In spite of this, by
verifying the vanishing curl condition (\ref{curl1}) one can
demonstrate the existence of a gradient flow.\footnote{In some cases a
direct integration turns out to be possible for an \emph{extremal}
ansatz, as in \cite{Hotta:2007wz,Gimon:2007mh}. One can readily check 
that the velocity field is irrotational in these examples.} For this 
reason we restrict ourselves to those theories that have a symmetric 
moduli space after timelike reduction, where we know that $f$ exists. 
It will therefore be convenient to now briefly review the relationship 
between black holes and geodesics on symmetric spaces.

\section{Black holes and geodesics}\label{s:FreeParticle}

Now we would like to examine the condition discussed in section
\ref{s:MultiScalar} for the generalised superpotential to exist. We
will consider a timelike reduction of the $D+1$-dimensional theory. We
showed that it suffices for the curl (\ref{curl1}) of a velocity field
on the scalar manifold of $D$ dimensions to vanish. In this section we
concentrate on theories for which the $D$-dimensional scalar manifold
is a symmetric space. For these theories we construct the velocity
field needed to investigate the curl-condition (\ref{curl1}). We begin
with explaining the timelike dimensional reduction to $D$ dimensions
and will then give the necessary background on symmetric spaces to
arrive at an expression for the velocity field $\dot \phi^A = \{\dot
\varphi, \dot \phi^a\}$.

\subsection{Timelike dimensional reduction}

The ansatz for stationary black holes can always be interpreted as the
ansatz for the dimensional reduction over time
\begin{align}
\de s^2_{D+1} &= -\e^{2\beta\varphi}(\de t - B^0)^2
+ \e^{2\alpha\varphi}\de s^2_D\,,\\
A^I &= \chi ^I(\de t - B^0) + B_m^I\de x^m\,,
\end{align}
where $B^0$ and $\varphi$ are the Kaluza--Klein (KK) vector and (KK)
dilaton, respectively. Normalisations are chosen in such a way that
the $(D,0)$-dimensional theory is in the Einstein frame and that
$\varphi$ is canonically normalised.

We will restrict to spherically symmetric solutions and truncate the
KK vector ($B^0=0$), since its presence would violate
staticity.\footnote{`Static' means that the spacetime admits a global,
nowhere zero, timelike hypersurface orthogonal Killing vector field. A
generalization are the `stationary' spacetimes, which admit a global,
nowhere zero timelike Killing vector field. In particular, stationary,
spherically symmetric spacetimes are static.} In $D+1=4$ we make an
exception: in $D=3$ the Taub-NUT vector $B^0$ can be dualised to a
scalar, $\tilde \chi^0$, which is part of the scalar manifold. We will 
make use of the group structure associated to this manifold and 
truncate the Taub-NUT scalar at the end of the calculation. In fact, 
when $D=3$ also $B^I$ can be dualised to axionic scalars
$\tilde \chi^I$. One can then verify that the kinetic terms of the 
axions $\chi^I$ and $\tilde \chi^I$ appear with the opposite sign 
\cite{Breitenlohner:1987dg}.

From (\ref{metric}) we make following ansatz for the dimensionally
reduced black hole (instanton)
\begin{equation}
\de s^2_D = \e^{2(D-1)A(\tau)}\de\tau^2+\e^{2A(\tau)}\de\Omega_{D-1}^2
\,,\qquad \tilde\phi^i = \tilde\phi^i(\tau)\,,\label{eq:D_metric}
\end{equation}
with $\tilde\phi^i$ denoting the scalars in the $D$-dimensional 
Euclidean theory
\begin{equation}
\tilde\phi^i = \{\phi^A, \chi^\alpha\}\,.
\end{equation}
The fields $\phi^A$ contain both the scalars of the $(D,1)$-
dimensional theory ($\phi^a$) and the KK dilaton $\varphi$, whereas 
the $\chi^\alpha$ are the axions $\chi^I$ (and $\tilde\chi^I$,
$\tilde\chi^0$ when $D=3$). The effective field equations, which arise 
by substituting ansatz \eqref{eq:D_metric} into the equations of 
motion, can be found by varying the following effective action (see 
e.g.~\cite{Janssen:2007rc})
\begin{equation}
S_\text{eff} =
\int\de\tau\Bigl((2\alpha^2)^{-1}(\dot{A}^2 + \e^{2(D-2)A}) 
- \tfrac12 \tilde G_{ij}\dot{\tilde\phi}^i \dot{\tilde\phi}^j\Bigr),
\label{D=3effective-action}
\end{equation}
where dots denote derivatives with respect to $\tau$. Note that we
reserve the symbol $\tilde G_{ij}$ for the moduli space metric in $D$
dimensions. This action has to be complemented by the Hamiltonian
constraint \cite{Janssen:2007rc}
\begin{equation}
\alpha^{-2}(\dot{A}^2 -\e^{2(D-2)A}) =
\tilde G_{ij}\dot{\tilde\phi}^i\dot{\tilde\phi}^j\equiv 2E\,,
\label{HamiltonianII}
\end{equation}
with $E$ a constant. Those $D$-dimensional solutions that lift to
extremal black holes in $(D,1)$ dimensions have flat $D$-dimensional
geometries, or equivalently $E=0$, which implies that the geodesic is
null
\begin{equation}\label{eq:null}
\tilde G_{ij}\dot{\tilde\phi}^i\dot{\tilde\phi}^j=0\,.
\end{equation}

In $D>3$ one can eliminate the $\chi^{\alpha}$ from the action, since 
the moduli space metric has the following properties
\begin{equation}
\tilde G_{\alpha A}=0\,,\qquad
\partial_{\alpha} \tilde G_{ij}=0\,,\qquad
\tilde G_{ab}=G_{ab}\,,\qquad
\tilde G_{\varphi\varphi}=1\,,\qquad
\tilde G_{\varphi a}=0\,.\label{properties}
\end{equation}
The first two identities can be derived from the fact that the shift
symmetries of the scalars $\phi^\alpha$ commute. In $D=3$ this is not
the case as the shift symmetries associated with electric and magnetic
charges $q_I,p^I$ no longer commute. With a slight abuse of notation,
we have
\begin{equation}
[p^I,q_J]=\Omega^I{}_J\,Q_\mathrm{T}\,,
\end{equation}
where $Q_\mathrm{T}$ is the NUT charge and $\Omega^I{}_J$ is a
symplectic invariant matrix. If the NUT charge is zero, the properties
of the moduli space metric (\ref{properties}) are also valid in $D=3$
upon truncation of the vectorial direction that corresponds to the NUT
charge. Henceforth we always restrict to solutions with a vanishing
NUT charge (thus, spherically symmetric in $D+1$ dimensions).

\subsection{Geodesics on symmetric spaces}

The assumption we shall make is that the target space in $D$
dimensions is a symmetric coset space $G/H$, where $G$ is a Lie group
and $H$ some subgroup subject to certain conditions that we shall
state below. In the theories we consider, such as various
supergravities in arbitrary dimensions,  the Lie algebra of $G$ is
always semi-simple. The condition that the target space is a symmetric
space is always valid for supergravity theories with more than eight
supercharges and is sometimes valid for theories with less
supersymmetry. Nevertheless, our analysis here is independent of any
supersymmetry considerations.

We will take $L$, an element of $G$, to be a coset representative. We
first define the group multiplication from the left, $L\rightarrow
gL$, $\forall g\in G$, and we let the local symmetry act from the
right $L\rightarrow Lh$, $\forall h \in H$. The definition of a coset
element of $G/H$ then implies that we identify $L$ and $Lh$. The Lie
algebras associated to $G$ and $H$ are denoted by $\mathfrak{g}$ and
$\mathfrak{h}$ respectively. The defining property of a symmetric
space $G/H$ is that there exists a Cartan decomposition
\begin{equation}
\mathfrak{g} = \mathfrak{h} + \mathfrak{f}\,,
\end{equation}
with respect to the Cartan automorphic involution $\theta$, such that
$\theta (\mathfrak{f})= - \mathfrak{f}$ and $\theta(\mathfrak{h}) = +
\mathfrak{h}$. From the Cartan involution we can construct the
symmetric coset matrix $M=LL^{\sharp}$, where $\sharp$ is the
generalised transpose, defined as
\begin{equation}
L^{\sharp}=\exp[-\theta(\log L)]\,.
\end{equation}
The matrix $M$ is invariant under $H$-transformations that act from
the right on $L$.  Under  $G$-transformations from the left, $M$
transforms as follows
\begin{equation}
M \rightarrow g M g^{\sharp}\,.
\end{equation}

With the aid of the matrix $M$ the line element on the space $G/H$
with coordinates $\tilde\phi^i$ can be written as
\begin{equation}
\de s^2 = \tilde G_{ij}\de\tilde\phi^i\de\tilde\phi^j
 = -\tfrac12\tr\bigl(\de M \de M^{-1}\bigr)\,.
\end{equation}
We expand the matrix valued one-form $M^{-1}\de M $ in the generators
$T_{\Lambda}$ of Lie algebra $\mathfrak{g}$ and in the basis
$\de\tilde\phi^i$ as follows
\begin{equation}
M^{-1}\de M
= (M^{-1}\partial_i M)^{\Lambda}\de\tilde\phi^i T_{\Lambda}
\equiv e^{\Lambda}_i\de\tilde\phi^i T_{\Lambda}\,,
\end{equation}
where we introduced the symbol $e^{\Lambda}_i$. This plays a role
similar to the vielbein on $G/H$, but one should note that
$e^{\Lambda}_i$ is not a square matrix, since $\Lambda = 1,\dotsc,
\dim G$ and $i = 1,\dotsc, \dim G/H$. In this language, the metric can
also be written as
\begin{equation}
\tilde G_{ij}=e^{\Lambda}_i\,\eta_{\Lambda\Sigma}\,e^{\Sigma}_j\,,
\qquad\eta_{\Lambda\Sigma}
=\tfrac{1}{2}\tr\left(T_{\Lambda}T_{\Sigma}\right).\label{eq:G-ee}
\end{equation}
In the above $\eta$ is proportional to the Cartan--Killing metric of
$\mathfrak{g}$ and is non-degenerate, since $\mathfrak{g}$ is
semi-simple in the theories we consider.

The metric is invariant under a $\emph{local}$ action of $H$ on $L$
from the right and under a \emph{global} action of $G$ on $L$ from the
left. The latter implies that $G$ is the isometry group of $G/H$, as
expected. The action for the geodesic curves on $G/H$ is then given by
\begin{equation}
S = \int\de\tau\,\,\tilde G_{ij}\dot{\tilde\phi}^i\dot{\tilde\phi}^j
= -\tfrac12 \int\de\tau\,
\tr\left(\tfrac{\de}{\de\tau}M \tfrac{\de}{\de\tau}(M^{-1})\right),
\label{eq:coset_action}
\end{equation}
where $\tau$ is an affine coordinate parametrising the geodesic curves
and the resulting equations of motion are
\begin{equation}\label{eq:scalarEOM}
\tfrac{\de}{\de \tau}(M^{-1}\tfrac{\de}{\de \tau}M)=0\qquad
\Rightarrow\qquad M^{-1}\tfrac{\de}{\de \tau}M=Q\,,
\end{equation}
with the matrix of Noether charges $Q$ being a constant matrix in some
representation of $\mathfrak{g}$. We now see that the geodesic
equations are indeed integrable and their general solution is
\begin{equation}
M(\tau)=M(0) \e^{Q\tau}\,.\label{SOLUTION}
\end{equation}
The affine velocity squared of the geodesic curve is (the dot stands
for ordinary matrix multiplication)
\begin{equation}\label{eq:affinev}
\tilde G_{ij}\dot{\tilde\phi}^i\dot{\tilde\phi}^j
= \tfrac{1}{2}\tr(Q\cdot Q)\,,
\end{equation}
and coincides with the Hamiltonian constraint \eqref{HamiltonianII}.

An integrable geodesic motion on an $n$-dimensional space is
characterised by $2n$ constants: the initial position and velocity of
the geodesic curve. In our case the geodesic motion on $G/H$ is thus
specified by $2(\dim G-\dim H)$ integration constants.  In equation
(\ref{SOLUTION}) $M(0)$ contains $(\dim G-\dim H)$ arbitrary constants
that correspond to the initial position and $Q$ corresponds to the
initial velocity, so we also expect $(\dim G-\dim H)$ arbitrary
constants there. This can be understood from the constraint
\begin{equation}\label{eq:GenInvCond}
M^{\sharp}(\tau)=M(\tau) \qquad \Rightarrow \qquad
\theta(Q)=-M(0)^{-1}QM(0)\,,
\end{equation}
which indeed reduces the number of arbitrary constants in $Q$ from
$\dim G$ to $(\dim G-\dim H)$.

The first-order equation (\ref{eq:scalarEOM}) can be written compactly
as $e^{\Lambda}_i\dot{\tilde\phi}^i=Q^{\Lambda}$ or equivalently,
\begin{equation}
\dot{\tilde\phi}^i = \tilde G^{ij} e_j^{\Sigma} \eta_{\Sigma\Lambda}
Q^{\Lambda}\,.\label{eq:fo_eqn-general}
\end{equation}
These are only $(\dim G-\dim H)$ equations. After substituting
\eqref{eq:fo_eqn-general} into eq.~\eqref{eq:scalarEOM}, the remaining
$\dim H$ components become non-differential equations. This shows the
power of \eqref{eq:scalarEOM}: we split the $\dim G$ differential
equations in $M^{-1}\tfrac{\de}{\de \tau}M=Q$ into $(\dim G-\dim H)$
first-order equations and $\dim H$ equations without any derivatives.
In the context of section \ref{s:Existence} these non-differential
equations are precisely what is needed to eliminate the additional
scalars resulting from dimensional reduction, so that we obtain
first-order equations in terms of the scalars in $D+1$ dimensions, as
in eq.~\eqref{velocity2}.

\section{The dilatonic black hole revisited}\label{single}

When the Einstein-dilaton-Maxwell action (\ref{eq:EMDaction}) has the
specific dilaton coupling $a=\pm\sqrt{3}$, a symmetry enhancement
takes place upon dimensional reduction over a (timelike or spacelike)
circle. This can be explained by the fact that (\ref{eq:EMDaction}) is
the action obtained from reducing five-dimensional gravity over a
spacelike circle and subsequent reduction should display at least the
$\GL(2,\Real)$ symmetry of the internal torus. Furthermore, if the 3d
vectors are dualised to scalars the $\GL(2,\Real)$-symmetry turns out
to be part of a larger $\SL(3,\Real)$-symmetry. The Euclidean 3d
action then describes gravity minimally coupled to the
$\SL(3,\Real)/\SO(2,1)$ sigma model with coordinates $\tilde\phi^i
=\{\phi^1,\phi^2,\chi^0,\chi^1,\chi^2\}$:
\begin{equation}
\begin{split}
\tilde G_{ij}\de\tilde\phi^i\de\tilde\phi^j
={}&(\de\phi^1)^2 + (\de\phi^2)^2
-\e^{-\sqrt{3}\phi^1+\phi^2}(\de\chi^0)^2+\e^{2\phi^2}(\de\chi^2)^2\\
&-\bigl[\e^{+\sqrt{3}\phi^1+\phi^2}-\e^{2\phi^2}(\chi^0)^2\bigr]
(\de\chi^1)^2 + 2\chi^0\e^{2\phi^2}\de\chi^1\de\chi^2\,.
\label{sigma}
\end{split}
\end{equation}
The details of the Kaluza--Klein reduction can be found in appendix
\ref{app:KK5to3}; for details on the $\SL(3,\Real)/\SO(2,1)$ sigma
model, see appendix \ref{SL3}. The scalars $\phi^1$ and $\phi^2$ are
both a linear combination of the black hole warp factor $\varphi$ and
the dilaton $\phi$. The scalars $\chi^0$ and $\chi^1$ are the electric
and magnetic potentials. $\chi^2$ comes from the dualisation of the KK
vector in the reduction from four to three dimensions and is hence
related to the NUT charge $Q_\mathrm{T}$ via
\begin{equation}\label{eq:NUT}
Q_\mathrm{T} \sim \dot{\chi}^2 + \chi^0\dot{\chi}^1\,.
\end{equation}
As explained before, a vanishing NUT charge leads to a truncated
target space, where $\chi^0$ and $\chi^1$ have a shift symmetry
\begin{equation}
\de s^2=(\de\phi^1)^2 + (\de\phi^2)^2 -
\e^{-\sqrt{3}\phi^1+\phi^2}(\de\chi^0)^2
-\e^{+\sqrt{3}\phi^1+\phi^2}(\de \chi^1)^2 \,.
\end{equation}
Upon eliminating $\de \chi^1$ and $\de\chi^2$ by their equations of
motion, one obtains the black hole potential for $\phi^1$ and
$\phi^2$.

Let us discuss the geodesic equations of motion for the full sigma
model (\ref{sigma}). The charge matrix $Q\in \mathfrak{sl}(3)$ that
specifies a geodesic solution contains eight arbitrary parameters $Q =
Q^\Lambda T_\Lambda$, where the eight generators of
$\mathfrak{sl}(3)$, denoted $T_{\Lambda}$, are given in the appendix,
eq.~\eqref{eq:T}. We will reduce the number of integration constants
to four. Firstly, we demand that the geodesic curve goes through the
origin which, using \eqref{eq:GenInvCond}, gives an involution
condition on $Q$
\begin{equation}\label{eq:InvCond}
Q = -\theta(Q)\,,
\end{equation}
and requires identifying
\begin{equation}\label{involution1}
Q^3=-Q^6\,,\qquad Q^4=-Q^7\,,\qquad Q^5=Q^8\,.
\end{equation}
This condition is gauge-equivalent to the most general expression for
$Q$. It amounts to fixing the $U(1)$-gauge transformation and the
boundary conditions for the black hole warp factor and dilaton at
spatial infinity. This specification is without loss of
generality.\footnote{We put $\phi(r\to\infty)=\varphi(r\to\infty)=0$.
This condition on the warp factor $\varphi$ can always be achieved by
a coordinate transformation. The condition for the dilaton cannot be
changed, but any other boundary value is equivalent upon a shift of
the dilaton and accordingly a compensating rescaling of magnetic and
electric charge.} Secondly, we restrict ourselves to solutions with a
vanishing NUT charge, which amounts to $Q^5=Q^8=0$, so that we are
left with
\begin{equation}
Q =Q^\Lambda T_\Lambda=\begin{pmatrix}
-\tfrac{1}{\sqrt{3}}Q^1-Q^2 & -Q^6 & 0\\
Q^6 & \tfrac{2}{\sqrt{3}}Q^1 & -Q^7\\
0& Q^7& -\tfrac{1}{\sqrt{3}}Q^1 +Q^2\end{pmatrix}.\label{eq:charge1}
\end{equation}
Thus we count four independent integration constants to describe the
dilatonic black hole solutions: the mass, the electric and magnetic
charge and the scalar charge. Upon demanding a regular horizon one can
write the scalar charge in terms of the three others
\cite{Gibbons:1982ih}, but we do not make that restriction here for
the sake of generality.

We now have sufficient information to construct the velocity vector
field $f^i(\phi^j) = \dot{\tilde\phi}^i$ for the charge configuration
(\ref{eq:charge1})
\begin{align}
f^{{\phi}^1}&=Q^1 + \tfrac{\sqrt{3}}{2}(Q^7 \chi^0 - Q^6
\chi^1)\,,\\
f^{{\phi}^2}&=Q^2 -\tfrac12( Q^7 \chi^0 + Q^6 \chi^1)\,,\\
f^{{\chi}^0}&=
Q^7 \e^{\sqrt{3}\phi^1 - \phi^2}\,,\label{eq:fo_chi0}\\
f^{{\chi}^1}&=
Q^6 \e^{-\sqrt{3}\phi^1- \phi^2}
\,,\label{eq:fo_chi1}\\
f^{{\chi}^2}&=-\chi^0 f^{\chi^1}\,.
\end{align}
Note that we have already used the component of the velocity field for
the Taub-NUT scalar ($\dot \chi^2 = f^{\chi^2}$) to eliminate $\dot
\chi^2$ from the other components of the velocity field via
eq.~\eqref{eq:NUT} with $Q_\mathrm{T}=0$. From the asymptotic
behaviour of the velocity field we can then identify the charges
$Q^\Lambda$: $Q^1$ is proportional to the ADM mass, $Q^2$ is
proportional to the dilaton charge, while $Q^6$ and $Q^7$ are equal to
the  magnetic and electric charge respectively.

Aside from the explicit expression for the velocity field, there is
more information in the \emph{eight} first-order equations
$M^{-1}\dot{M}=Q$. The velocity field uses five out of these eight.
The remaining three equations are non-differential and we call them
\emph{constraint equations}:
\begin{align}
0&=Q^6 + e^{\sqrt{3} \phi^1+\phi^2} \bigl[\bigl(\sqrt{3}
Q^1+Q^2\bigr) \chi^1-Q^6 \bigl(1+(\chi^1)^2\bigr)-Q^7
\chi^2\bigr]\,,\\
0&=Q^7+e^{-\sqrt{3} \phi^1+\phi^2} \bigl[\bigl(-\sqrt{3}
Q^1+Q^2\bigr)\chi^0 -Q^7 \bigl(1+(\chi^0)^2\bigr) +Q^6( \chi^2
+\chi^0\chi^1)\bigr]\,,\\
\begin{split}
0&= 2 Q^2(2\chi^2 + \chi^0\chi^1)
-Q^6\bigl[\bigl(1+e^{-\sqrt{3} \phi^1-\phi^2}\bigr)\chi^0 + \chi^1
\chi^2\bigr]\\
&\quad+Q^7 \bigl[\bigl(1+e^{\sqrt{3} \phi^1-\phi^2}\bigr)
\chi^1-\chi^0 (\chi^2+\chi^0 \chi^1)\bigr]\,.
\end{split}
\end{align}
Note that we already used the constraint equations to simplify the
first order derivatives of $\chi^0$ and $\chi^1$: (\ref{eq:fo_chi0})
and (\ref{eq:fo_chi1}). The constraint equations (at least
theoretically) enable one to extract the functional dependence of the
$\chi^{\alpha}$ on the $\phi^1$ and $\phi^2$, such that we can write
$f^{\phi^1}(\phi,\chi)$ and $f^{\phi^2}(\phi,\chi)$ purely in terms of
$\phi^1$ and $\phi^2$:
\begin{equation}
f^{\phi^1}(\phi)\equiv f^{\phi^1}[\phi,\chi(\phi)]\,,\qquad
f^{\phi^2}(\phi)\equiv f^{\phi^2}[\phi,\chi(\phi)]\,.
\end{equation}
The condition for the existence of a first-order gradient flow then
becomes
\begin{equation}\label{curl2}
\partial_{[\phi^1}f_{\phi^2]} = -\tfrac{1}{4}Q^7
(\sqrt{3}\partial_{\phi^2}+\partial_{\phi^1})\chi^0(\phi^1,\phi^2)
 + \tfrac14Q^6(\sqrt{3}\partial_{\phi^2}
 - \partial_{\phi^1})\chi^1(\phi^1,\phi^2)=0\,,
\end{equation}
and we can evaluate under which conditions on the charges $Q^\Lambda$
the expression (\ref{curl2}) holds.

In principle, we have three constraint equations at our disposal to
eliminate the three axions $\chi^\alpha(\phi^1,\phi^2),\alpha=0,1,2$,
but in practice this would require solving relatively complicated
non-linear simultaneous equations, which is not straightforward.
Fortunately, the curl (\ref{curl2}) requires only a knowledge of the
derivatives of the axions with respect to the dilatons, i.e.~the
Jacobian matrix $[J^{\alpha}_A](\phi)\equiv \partial_{\phi^A}
\chi^{\alpha}$, $\alpha=0,1$. It turns out that the inverse Jacobian 
matrix $[J^A_{\alpha}](\chi) \equiv \partial_{\chi^{\alpha}}\phi^A$ is 
easily computable using the constraint equations. If we then use
\begin{equation}
[J^{\alpha}_A](\phi(\chi))=[J_{\alpha}^A]^{-1}(\chi)\,,
\end{equation}
where the inverse is with respect to the whole matrix, we can evaluate
the curl in terms of the fields $\chi^0, \chi^1$. An explicit
calculation shows that the curl vanishes. We thus conclude that
\emph{when $a=\sqrt{3}$, all the dilatonic black holes with arbitrary
mass, electric, magnetic and scalar charge possess a generalised
superpotential.}

We have not attempted the construction of the generalised
superpotential for arbitrary solutions with $a = \sqrt{3}$, but rather
only for extremal cases. Then one can use the factorisation property
\eqref{factorization} of the superpotential to deduce $W$ as a
function of $\tau$ from the flow equation \eqref{FLOW'II} with the
help of the known explicit solution \cite{Gibbons:1982ih} and invert
$\phi(\tau)$ to obtain $W(\phi)$. Even in this simplified setting the
result is very long compared to the $a=1$ case and not illuminating,
we therefore refrain from quoting it here.

\section{Kaluza--Klein black hole in five dimensions}\label{multiple}

Let us now consider  black holes carried by multiple scalars and
vectors. In $D+1=5$ there is an example for which we can use the same
hidden symmetry as for the KK dilatonic black hole, namely
$\SL(3,\Real)$. This theory is obtained by reducing 7d gravity on a
two-torus. This gives a 5d theory with two vectors, and three scalars:
an axion-dilaton system and an extra dilaton $\tilde\varphi$
\begin{equation}\label{eq:KKBHaction}
S=\int\de^5
x\sqrt{|g|}\Bigl(\mathcal{R}-\tfrac{1}{2}
(\partial\tilde\varphi)^2+\tfrac { 1 } { 4 }\tr(\partial
K\partial K^{-1})
-\tfrac{1}{4}\e^{\sqrt{\frac{5}{3}}\tilde\varphi}K_{mn}F^m
F^n\Bigr) ,
\end{equation}
where the matrix $K$ defines the $\SL(2,\mathbb{R})$ axion-dilaton
system. Details on this action and subsequent reduction to 4
dimensions can be found in appendix \ref{app:KK7to5}.

This Lagrangian is a consistent truncation of maximal and half-maximal
supergravity in $D+1=5$. Upon reduction over time one obtains
four-dimensional Euclidean gravity coupled to a set of scalars that
span the coset $\SO(1,1)\times \SL(3,\Real)/\SO(2,1)$. The dynamics of
the decoupled scalar (the $\SO(1,1)$ part) is trivial and the
$\SL(3,\Real)/\SO(2,1)$ part differs from the previous example only in
that this coset has a different $\SO(2,1)$ isotropy group embedded in
$\SL(3,\Real)$. The effect of this is purely a matter of signs, as can
be seen in the metric on the moduli space (neglecting the decoupled
scalar, see appendix \ref{app:KK7to5})
\begin{equation}
\begin{split}
\tilde G_{ij}\de\tilde\phi^i\de\tilde\phi^j
={}&(\de\phi^1)^2 + (\de\phi^2)^2 +
\e^{-\sqrt{3}\phi^1+\phi^2}(\de\chi^0)^2 - \e^{2\phi^2}(\de\chi^2)^2\\
&-\bigl[\e^{+\sqrt{3}\phi^1+\phi^2} + \e^{2\phi^2}(\chi^0)^2\bigr]
(\de\chi^1)^2 - 2\chi^0 \e^{2\phi^2} \de\chi^1\de\chi^2\,.
\label{sigma2}
\end{split}
\end{equation}
This sigma model can be obtained from (\ref{sigma}) through the
analytic continuation
\begin{equation}
\chi^0\rightarrow \im\chi^0\,,\qquad \chi^2 \rightarrow \im\chi^2\,.
\end{equation}
The representative $\tilde L$ of the full coset $\SO(1,1)\times
\SL(3,\Real)/\SO(2,1)$ is then given by
\begin{equation}
\tilde L = \e^{\phi^0/\sqrt{6}} L\,,
\end{equation}
where $L$ is the $\SL(3,\Real)/\SO(2,1)$ coset representative
\eqref{L} and the decoupled scalar $\phi^0$ is related to
$\tilde\varphi$ of \eqref{eq:KKBHaction} by eq.~\eqref{eq:phi0}.

We will again assume that the charge matrix describes only the
geodesics that go through the origin. As before we can justify this
restriction by proper field redefinitions and coordinate
transformations of the general solution. The Cartan involution
condition \eqref{eq:InvCond} implies (cf.~\eqref{involution1})
\begin{equation}
Q^3 = -Q^6, \qquad Q^4 = Q^7,  \qquad Q^5 = -Q^8\,,
\end{equation}
so that
\begin{equation}
Q = Q^\Lambda T_\Lambda =
\begin{pmatrix}
Q^0-\frac{Q^1}{\sqrt{3}}-Q^2 & -Q^6 & -Q^8 \\
Q^6 & Q^0+\frac{2 Q^1}{\sqrt{3}} & Q^7 \\
Q^8 & Q^7 & Q^0-\frac{Q^1}{\sqrt{3}}+Q^2
\end{pmatrix},
\end{equation}
where now $\Lambda = 0,\dotsc,8$, $T_0$ is the three-dimensional
identity matrix generating the decoupled $\SO(1,1)$ part and the
remaining generators are, as previously, given by \eqref{eq:T}. The
parameters $Q^6$ and $Q^8$ can be identified with the electric charges
in $D+1=5$.

To obtain the first-order velocity field for the effective action with
the black hole potential one needs to eliminate $\chi^1$ and $\chi^2$
in terms of the remaining scalars using the constraint equations,
which can be concisely written as
\begin{align}
\dot\chi^0 &=\e^{+\sqrt{3}\phi^1-\phi^2}\bigl(Q^7-Q^8\chi^1\bigr)\,,\\
\dot\chi^1 &=\e^{-\sqrt{3}\phi^1-\phi^2}\bigl(Q^6-Q^8\chi^0\bigr)\,,\\
\dot\chi^2 &=\e^{-2\phi^2}Q^8 - \chi^0\dot\chi^1 = \e^{-2\phi^2}Q^8
-\e^{-\sqrt{3}\phi^1-\phi^2}\chi^0\bigl(Q^6 - Q^8\chi^0\bigr)\,,
\end{align}
where the left-hand side is understood to be expressed by
eq.~\eqref{eq:fo_eqn-general} and does not contain derivatives. Unlike
in the dilatonic black hole example, there are more constraints than
variables to eliminate, unless specific choices for the charges make
fewer of them independent. Using different combinations of constraint
equations to eliminate $\chi^1$ and $\chi^2$ leads to different
velocity fields in five dimensions. Although they become equivalent
upon using the Hamiltonian constraint (which is exactly the remaining
constraint equation), the expression for the curl is not unique. One
preferred form should however distinguish itself, namely that not
containing second-order integration constants. Finding such a
combination of constraint equations is a technically complex task, as
it involves relaxing the boundary conditions ($M(0)=1$) in order to
distinguish first- and second-order integration
constants.\footnote{The first-order integration constants in $Q$ and
the second-order integration constants in $M(0)$ are intertwined
through the involution condition (\ref{eq:GenInvCond}) making it
difficult to distinguish them in the coset matrix formalism.} For this
reason we have not pursued it further.

Regardless of which combination of constraint equations should serve
to eliminate the extraneous scalars $\chi^1$ and $\chi^2$, the
resulting expression for the curl in five dimensions will be
non-trivial and will not involve $Q^0$, hence the condition for the
curl to vanish is independent of extremality, which in turn amounts to
(cf.~remarks preceding equations \eqref{eq:null} and
\eqref{eq:affinev})
\begin{equation}
\tr(Q\cdot Q) = 3 (Q^0)^2 + 2\left[(Q^1)^2 + (Q^2)^2 - (Q^6)^2 +
(Q^7)^2 - (Q^8)^2\right] = 0\,.
\end{equation}
We conclude that among  both extremal and non-extremal solutions there
exist examples that admit a generalised superpotential, but also
examples that do not.

\section{Discussion}\label{s:discussion}

\subsection{Summary of results}

For theories of gravity coupled to neutral scalar fields and Abelian
vector fields, we have presented the most general form of first-order
flow equations consistent with rewriting the effective action as a sum
(or difference) of squares. The derivatives of the scalars with
respect to the radial parameter are given by the gradient of a
generalised superpotential on the scalar manifold (equations
\eqref{FLOWII}, \eqref{FLOWIII}). The generalised superpotential is
related to the black hole potential by eq.~\eqref{Y-equation}. The
above gradient flow equations are equally applicable to extremal
(whether supersymmetric or not) as well as non-extremal black holes
(necessarily non-supersymmetric). They naturally encompass previously
known partial results, although they differ from the form conjectured
in \cite{Andrianopoli:2007gt}.

We considered theories with scalar manifolds which become symmetric
spaces after a timelike dimensional reduction and produced a method to
verify when a generalised superpotential exists. We have provided
examples of extremal and non-extremal solutions with a generalised
superpotential, but also shown that it is possible to find solutions
(including extremal ones) for which one cannot exist.

Let us now discuss the examples in which we obtained the above results
in more detail. We have applied our formalism to a dilatonic black
hole in four dimensions (one scalar field) and a Kaluza--Klein black
hole in five dimensions (multiple scalars). For the dilatonic black
hole with the dilaton coupling $a=1$ we were able to show by direct
integration that the generalised flow equations exist in all
situations. When $a=\sqrt{3}$ we were able to show the same, using
group-theoretical tools to integrate the second-order equations of
motion to first-order equations. For all other values of $a$ we
derived the existence of a fake superpotential in the \emph{extremal}
case, using the argument applied for single scalar domain walls
\cite{Skenderis:2006jq}. The existence of generalised flow equations
for non-extremal black holes with arbitrary dilaton coupling is not
known to us. Although the $a=1$ case was easy to integrate by hand,
this is also an example for which we could have constructed the flow
using group theory. The reason is that this case is embeddable in an
$\cN=4$ action, which has a symmetric moduli space after timelike
reduction: $\SO(8,8+n)/[\SO(6,2)\times \SO(6+n,2)]$ (see for instance
\cite{Bergshoeff:2008be}). The investigation of the Kaluza--Klein
black hole in five dimensions, in turn, demonstrated that for both
extremal and non-extremal solutions, there are cases where a
generalised superpotential exists and where it does not, depending on
the values of the scalar and vector charges.

The same techniques can be applied to more complicated examples.
Possible further work might include exploring other non-extremal
cases, a natural candidate being the $STU$ or the $T^3$ model in
$\cN=2$ supergravity. It would be most interesting to see whether
there exists a closed form of the generalised superpotential,
universally valid for all cosets. It might be also useful to
investigate whether the vanishing curl condition \eqref{curl1} could
be given a physical interpretation in terms of other black hole
properties. A broader problem that suggests itself for study is one of
the existence of a generalised superpotential in theories whose scalar
manifolds are not symmetric spaces after a timelike dimensional
reduction.

\subsection{Comparison with domain walls}

It has been noticed in the literature that black hole effective
actions are very similar to domain-wall (and cosmology) effective
actions \cite{Ceresole:2007wx, Behrndt:2001qa}. Roughly speaking, the
only difference is in the expression for the potential in terms of the
`superpotential'
\begin{equation}\label{ksi}
V \sim W^2 + \xi (\partial W)^2\,,
\end{equation}
where the constant $\xi$ is positive for black holes and negative for
domain walls (and cosmologies). The precise value for $\xi$ depends on
the dimension. We would like to point out that this is more then just
an analogy as spherically symmetric black holes are nothing but domain
walls seen from a $1+1$ dimensional point of view. This is consistent
with (\ref{ksi}) since, in the domain wall case, the usual formula for
$\xi$ diverges when $D+1=2$ and instead one has to use the expression
for $\xi$ for black hole effective potentials. This exact
correspondence between spherically symmetric black holes and domain
walls in two dimensions comes about in the same way as the
correspondence between domain wall solutions of gauged supergravity
theories in $2<D+1<10$ and $p$-brane (and M-brane) solutions in
$D+1=10$ (and $D+1=11$) \cite{Cvetic:1999xx, Bergshoeff:2004nq}. There
the correspondence was obtained by considering spherical flux
reductions of type II theories to some lower-dimensional gauged
supergravity. The domain-wall solutions of the latter theory lift to
various (distributions of) $p$-brane solutions, whose metrics possess
the required spherical symmetry. Similarly, the construction of the
black hole effective potential can effectively be seen as an $S^2$
reduction of 4d ungauged supergravity to a 2d gauged supergravity. The
charges that appear in the black hole effective potential correspond
to the flux parameters of the `flux compactification'.

As in the black hole case, it was also appreciated that the effective
action for domain walls can be described both as a free particle
action and as a particle subject to a potential. In the case of
cosmological solutions this is known as `cosmology as a geodesic
motion' \cite{Townsend:2004zp} (see also \cite{Bergshoeff:2008zz}). By
virtue of the `domain wall/cosmology correspondence'
\cite{Skenderis:2006jq}, the same principle applies to domain wall
solutions. It is in this sense that the existence of two different
types of effective actions for black holes can be understood: it is
the same as `domain walls as a geodesic motion'
\cite{Townsend:2004zp}, but applied to $D=2$ domain walls. Finally, in
a pure mathematical context, this correspondence between particle
actions with a force field and an associated action of a free particle
in an enlarged target space is what underlies the way the
integrability of Toda--Liouville equations is linked to the
integrability of the geodesic equations on symmetric spaces
\cite{Ferreira:1984bi}.

\acknowledgments

We are grateful to Wissam Chemissany, Iwein De Baetselier, Jan De
Rydt, Mario Trigiante, Antoine Van Proeyen and Dennis Westra for
helpful discussions. We also thank Gianguido Dall'Agata for useful
comments on the first version of this work. P.S. is supported by the
German Science Foundation (DFG). T.V.R. would like to thank the Junta
de Andaluc\'ia and the University of Oviedo for financial support. The
work of J.P. and B.V. has been supported in part by the European
Community's Human Potential Programme under contract
MRTN-CT-2004-005104 `Constituents, fundamental forces and symmetries
of the universe', in part by the FWO-Vlaanderen, project G.0235.05 and
in part by the Federal Office for Scientific, Technical and Cultural
Affairs through the `Interuniversity Attraction Poles Programme --
Belgian Science Policy' P6/11-P\@. B.V. is aspirant FWO-Vlaanderen.

\appendix
\section{Dimensional reductions}

\subsection{From $D+2=4+1$ to $D+1=3+1$ to $D=3$}\label{app:KK5to3}

The reduction of pure gravity in $4+1$ dimensions on a spacelike
circle leads to the Einstein-Maxwell-dilaton action
(\ref{eq:EMDaction}) with $a=\sqrt{3}$. When this is further reduced
over a timelike circle through the following reduction
ansatz\footnote{Note that in equation (A.1) $\phi^2$ is a scalar with
an upper index and not a square.}
\begin{align}
\de s_4^2 &=\e^{\phi^2}\de s_3^2 -\e^{-\phi^2}(\de t- B_t)^2\,,\\
A &= B_z - \chi^0 (\de t- B_t)\,, \qquad \phi^1=-\phi\,,
\end{align}
where $\phi$ is the dilaton in the four-dimensional theory, the
resulting 3d Euclidean action is given by
\begin{equation}\label{eq:dilatonicS3}
S_3=\int\sqrt{g}\Bigl(\mathcal{R}_3
-\tfrac{1}{2}\partial\varphi\partial\varphi
+\tfrac{1}{4}\tr(\partial K \partial K^{-1})
-\tfrac{1}{4}\e^{-\sqrt{3}\varphi} K_{mn}F^m F^n\Bigr),
\end{equation}
where $m,n=t,z$ and
\begin{align}
\varphi&=\tfrac{1}{2}\phi^1 + \tfrac{\sqrt{3}}{2}\phi^2\,,\\
K_{tt}&=-\e^{\tfrac{\sqrt{3}}{2}\phi^1-\tfrac{1}{2}\phi^2}
+(\chi^0)^2\e^{-\tfrac{\sqrt{3}}{2}\phi^1+\tfrac{1}{2}\phi^2}\,,\\
K_{zz}&=\e^{-\tfrac{\sqrt{3}}{2}\phi^1+\tfrac{1}{2}\phi^2}\,,\\
K_{tz}&=\e^{-\tfrac{\sqrt{3}}{2}\phi^1+\tfrac{1}{2}\phi^2}\chi^0\,.
\end{align}
If we dualise the 3d vectors to scalars via
\begin{equation}
\e^{-\sqrt{3}\varphi} K_{zn}\Hodge F^n\equiv\de \chi^1\,,\qquad
\e^{-\sqrt{3}\varphi} K_{tn}\Hodge F^n\equiv\de \chi^2\,,
\end{equation}
we find the sigma model (\ref{sigma}).

\subsection{From $D+3=6+1$ to $D+1=4+1$ to $D=4$}\label{app:KK7to5}

If we reduce pure 7d gravity, given by the action:
\begin{equation}
  \int \de^7 x \sqrt{|g_7|} \mathcal{R}_7\,,
\end{equation}
over a spacelike two-torus (with coordinates $y^m$, $m = 1,2$) via
\begin{equation}
\de s^2_7= \e^{2\alpha\tilde\varphi}\de s^2_5 + \e^{2 \beta
\tilde\varphi} K_{mn}(\de y^m + A^{(m)})(\de y^n + A^{(n)})\,,\quad
\alpha = -\sqrt{\tfrac 1{15}}\,,\quad
\beta = \tfrac 12\sqrt{\tfrac35}\,,
\end{equation}
we find the 5d action
\begin{equation}
S= \int \de^5 x \sqrt{|g_5|} \Bigl(\mathcal{R}_5 -\tfrac12 \partial
\tilde\varphi
\partial \tilde\varphi+\tfrac14 \tr(\partial K^{-1} \partial
K)-\tfrac14\e^{\sqrt{\tfrac53}\tilde\varphi} K_{mn} F^m
F^n\Bigr)\,.
\end{equation}
The $\SL(2,\mathbb{R})$ matrix $K$ parameterises the deformations of
the torus through the two scalars $\varphi^0$ and $\chi^0$:
\begin{equation}
  K = \e^{-\varphi^0}\begin{pmatrix}
    \e^{2\varphi^0} + (\chi^0)^2
    &\chi^0 \\ \chi^0 &1
  \end{pmatrix}\,.
\end{equation}
In five dimensions, this gives rise to  an $\SO(1,1) \times
\frac{\SL(2,\mathbb{R})}{\SO(2)}$ sigma model, where the decoupled
$\SO(1,1)$ is parameterised by $\varphi$, as can be seen from the
action.

A subsequent timelike reduction to $D=4+0$ via the ansatz (we truncate
the 4d vectors)
\begin{align}
\de s_5^2 &= -\e^{-2 \varphi^1}\de t^2 +\e^{\varphi^1}\de s^2_4\,,\\
A^{(1)}& =\chi^1\de t\,,\qquad A^{(2)} =\chi^2\de t\,,
\end{align}
gives the 4d action
\begin{equation}
\begin{split}
S={}&\int \de^4 x \sqrt{g_4}\Bigl[\mathcal{R}_4-\tfrac12 \partial
\tilde\varphi\partial \tilde\varphi - \tfrac12 \partial
\varphi^0\partial\varphi^0 -\tfrac32\partial\varphi^1\partial\varphi^1
-\tfrac12\e^{-2\varphi^0}\partial\chi^0\partial\chi^0\\
&+\tfrac12\e^{\sqrt{\tfrac53}\tilde\varphi + 2\varphi^1 -\varphi^0}
\Bigl(\bigl(\e^{2\varphi^0}+(\chi^0)^2\bigr)\partial
\chi^1\partial\chi^1
+2\chi^0\partial \chi^1\partial\chi^2+\partial \chi^2
\partial\chi^2 \Bigr)\Bigr]\,.
\end{split}
\end{equation}
This action describes an $\SO(1,1)\times
\frac{\SL(3,\mathbb{R})}{\SO(2,1)}$ sigma model coupled to gravity.
Written in this way, it is not evident how to decouple the $\SO(1,1)$
part. To obtain the form (\ref{sigma2}), we further have to perform
the following rotation of the dilatons
$\{\tilde\varphi,\varphi^0,\varphi^1\}\to\{\phi^0,\phi^1,
\phi^2\}$:
\begin{align}
  \phi^0 &= -\tfrac2{3}\tilde\varphi
+\sqrt{\tfrac53}\varphi^1\,, \label{eq:phi0}\\
  \phi^1 &= \tfrac1{2\sqrt{3}}\Bigl(\sqrt{\tfrac53}\tilde\varphi+3
\varphi^0+2\varphi^1\Bigr),\\
  \phi^2 &= \tfrac12 \Bigl(\sqrt{\tfrac53}\tilde\varphi
-\varphi^0+2\varphi^1\Bigr).
\end{align}
The action now becomes:
\begin{equation}\label{eq:SKKin4}
\begin{split}
  S={}&\int\de^4 x \sqrt{g_4}\Big{(}\mathcal{R}_4-\tfrac12 \partial
\phi^0
\partial \phi^0 -\tfrac12\partial\phi^1\partial\phi^1 -\tfrac12
\partial\phi^2\partial\phi^2 -\tfrac12
\e^{-\sqrt{3}\phi^1+\phi^2}\partial\chi^0\partial\chi^0\\
&+\tfrac12 \e^{2\phi^2}\partial \chi^2\partial \chi^2
+\tfrac12\bigl(\e^{\sqrt{3}\phi^1+\phi^2}
+\e^{2\phi^2}(\chi^0)^2\bigr)\partial
\chi^1\partial \chi^1 +\chi^0 \e^{2\phi^2}
\partial\chi^1\partial\chi^2\Big{)}\,.
\end{split}
\end{equation}

The scalar $\phi^0$ describes the $\SO(1,1)$ part, while the others
parameterise a $\frac{\SL(3,\mathbb{R})}{\SO(2,1)}$ sigma model.

\section{The $\frac{\SL(3,\Real)}{\SO(2,1)}$ sigma model}\label{SL3}

We define the $\SL(3,\Real)/\SO(2,1)$ coset element in the Borel gauge
\begin{equation} L=\exp(\chi^1 E_{12})\exp(\chi^0 E_{23})
\exp(\chi^2 E_{13})
\exp(\tfrac{1}{2}\phi^1H_0+\tfrac{1}{2}\phi^2H_2)\,,
\end{equation}
where $H_{1}$ and $H_{2}$ are the Cartan generators of
$\mathfrak{sl}(3)$ and the $E_{\alpha}$ are the three positive root
generators. In here we use the fundamental representation of
$\mathfrak{sl}(3)$ and choose the following basis for the generators
\begin{equation}
H_0 = \frac{1}{\sqrt{3}} \begin{pmatrix} -1 & 0 & 0 \\
0 & 2 & 0 \\
0 & 0 & -1
\end{pmatrix},
\qquad H_1 = \begin{pmatrix} -1 & 0 & 0 \\
0 & 0 & 0 \\
0 & 0 & 1
\end{pmatrix},\end{equation}
and the three positive step operators
\begin{equation}\label{Es}
E_{12} = \begin{pmatrix} 0 & 1 & 0 \\
0 & 0 & 0 \\
0 & 0 & 0
\end{pmatrix},
\qquad E_{23} = \begin{pmatrix} 0 & 0 & 0 \\
0 & 0 & 1 \\
0 & 0 & 0
\end{pmatrix},
\qquad E_{13} = \begin{pmatrix} 0 & 0 & 1 \\
0 & 0 & 0 \\
0 & 0 & 0
\end{pmatrix}.
\end{equation}
The generators $T_{\Lambda}$, $\Lambda=1,\ldots, 8$, of $\SL(3,\Real)$
are given by
\begin{equation}\label{eq:T}
T_{\Lambda}=\{H_0, H_1, E_{12}, E_{23}, E_{13}, E_{12}^\mathrm{T},
E_{23}^\mathrm{T},E_{13}^\mathrm{T} \}\,.
\end{equation}

Then the coset element is explicitly given by
\begin{equation}\label{L}
L=\begin{pmatrix} \e^{-\tfrac{1}{2\sqrt{3}}\phi^1-\tfrac{1}{2}\phi^2}
& \e^{\tfrac{\phi^1}{\sqrt{3}}}\chi^1 &
\e^{-\tfrac{\phi^1}{2\sqrt{3}}+\tfrac{\phi^2}{2}}(\chi^0\chi^1
+\chi^2 )\\
0 & \e^{\tfrac{\phi^1}{\sqrt{3}}} &
\e^{-\tfrac{\phi^1}{2\sqrt{3}}+\tfrac{\phi^2}{2}}\chi^0 \\
0 & 0 & \e^{-\tfrac{\phi^1}{2\sqrt{3}}+\tfrac{\phi^2}{2}}
\end{pmatrix}.
\end{equation}
To find the metric on the coset we define the symmetric coset matrix
$M$ via $M=L\eta L^\mathrm{T}$ where $\eta$ is the matrix whose
stabiliser defines the specific isotopy group $\SO(2,1)$ of the coset.
To reproduce the sigma model (\ref{sigma}) we choose
\begin{equation}
\eta=\operatorname{diag}(+1,-1,+1)\,,
\end{equation}
whereas the other sigma model (\ref{sigma2}) has another $\SO(2,1)$
defined by
\begin{equation}
\eta=\operatorname{diag}(-1,+1,+1)\,.
\end{equation}
The metric that is then defined by $\de s^2=-\tfrac{1}{2}\tr(\de M\de
M^{-1})$ and the Cartan involution for a matrix
$A\in\mathfrak{sl}(3,\Real)$ is
\begin{equation}
\theta(A) = -\eta A^\mathrm{T}\eta\,.
\end{equation}

\bibliographystyle{utphys}
\bibliography{flow}

\end{document}